\documentclass[11pt]{llncs}

\begin{document}
\title{Eigenvalue Formulation of Quantum Mechanics Near
Closed Timelike Curves}

\author{Z. Gedik}

\institute{Faculty of Engineering and Natural Sciences, Sabanci
University, 34956 Tuzla, Istanbul, Turkey}

\maketitle
\begin{abstract}
Einstein's field equations of gravitation are known to admit closed
timelike curve (CTC) solutions. Deutsch approached the problem from
the quantum information point of view and proposed a
self-consistency condition. In this work, the Deutsch equation is
formulated as an eigenvalue problem. The disappearance of
entanglement between two qubits in an Einstein-Podolsky-Rosen (EPR)
state near a CTC is demonstrated. The method is utilized to analyze
the discontinuous evolution of two chronology respecting (CR) qubits
near a CTC.
\end{abstract}

\section{Introduction}

G\"{o}del pointed out the possibility of closed timelike curve (CTC)
solutions for the Einstein's field equations of gravitation
\cite{Godel}. Several models and calculations gave evidence for
existence of CTCs
\cite{Carter,Tipler,Bonnor,Morris,Frolov,Kim,Gott,Hawking,Ori}.
Because CTCs allow a time traveling particle to go back in time and
interact with its own past, one can end up with paradoxes. Deutsch
analyzed the problem from the quantum information point of view and
proposed a self-consistency condition involving the density matrices
of the chronology respecting (CR) and CTC qubits undergoing a
unitary interaction \cite{Deutsch}. Deutsch showed that the
self-consistency equation implies a non-unitary and nonlinear
evolution for the CR components. Ralph introduced a toy model to
show that unitarity can be recovered \cite{Ralph1}.

Consequences of existence of CTCs for quantum computation have been
examined. Brun argued that using CTCs, composite numbers can be
factorized efficiently with the help of a classical computer
\cite{Brun1}. Bacon demonstrated that nonlinearity of Deutsch-like
evolution can be used to efficiently solve computational problems by
reducing NP-hard problems to P \cite{Bacon}. It has also been
claimed that an observer with access to CTCs can perfectly
distinguish nonorthogonal quantum states \cite{Brun2}. However,
there are both counter and supporting arguments for this claim
\cite{Bennett,Ralph2}.

An alternative formulation of the quantum mechanics near CTCs is via
post-selection \cite{Lloyd1,Lloyd2}. Lloyd \emph{et al.}, developed
a model, which is physically inequivalent to Deutsch's, based on
combining quantum teleportation with post-selection. Unlike
Deutsch's mixed state generating CTCs, post-selected CTCs always
send pure states to pure states, and hence, they do not create
entropy.

Pati \emph{et al.}, showed that in contrast to ordinary quantum
mechanics where any mixed state can be purified by bringing an
ancillary system, the states of quantum systems traveling in CTCs
cannot be purified \cite{Pati}. In other words, the CTC system
interacting with the CR system cannot be viewed as a part of a
larger CTC system  in a pure entangled state in an enlarged Hilbert
space. A practical consequence of this result is that in general it
is not possible to simulate CTC qubits satisfying the Deutsch
equation in laboratory, because arbitrary CTC states cannot be
constructed in a consistent way.

In this study, after introducing an eigenvalue formulation of the
Deutsch equation, the method is used to demonstrate that the
entanglement between two qubits in Einstein-Podolsky-Rosen (EPR)
state disappears when one of the two qubits interacts with a CTC
qubit. The same formulation is utilized to examine the discontinuous
evolution of CR qubit states, first pointed out by DeJonghe \emph{et
al.} \cite{DeJonghe}.

\section{The Deutsch Equation}

Deutsch's model involves a unitary interaction $U$ of a CR system
with another system that traverses a CTC. The equation is a result
of the kinematic self consistency condition, which implies that the
density matrix of the CTC system after it interacts with the CR
system is the same as that of the CTC system before the interaction.
For each state of the CR register described by the density matrix
$\rho_{CR}$, CTC register is postulated to find a fixed point such
that
\begin{equation}\label{de}
Tr_{CR}\left(U\rho_{CR}\otimes\rho_{CTC}U^\dagger\right)=\rho_{CTC},
\end{equation}
where $\otimes$ is tensor product and $U^\dagger$ is the hermitian
conjugate of $U$. Here, it is assumed that CR and CTC system are
initially in a separable state. Deutsch showed that there is always
a density matrix $\rho_{CTC}$ which satisfies the above
self-consistency condition by using the compactness of the space of
density operators. In fact, there are in general many solutions, and
one might need additional assumptions to decide which $\rho_{CTC}$
to choose \cite{Deutsch,Bacon,DeJonghe,Politzer}. One of the
immediate consequences of the equation is that the universe (i.e.,
the space of CR registers) may evolve from a pure state to a mixed
state. Nonlinear and noncontractive evolutions are among the other
unusual effects \cite{Deutsch,Cassidy,Bacon}.

Equation~(\ref{de}) implies that $\rho_{CTC}$, and hence, its von
Neumann entropy
\begin{equation}
S\left(\rho_{CTC}\right)=-Tr\left(\rho_{CTC}\ln\rho_{CTC}\right)
\end{equation}
remains the same after CR and CTC systems interact. Therefore, the
entanglement of $\rho_{CTC}$ with the rest of the the enlarged CTC
Hilbert space would remain the same for any local interaction if it
were possible to distill $\rho_{CTC}$, which runs counter to
observation of Pati \emph{et al.} \cite{Pati}. However, for
arbitrary local operations, the entropy does not remain constant but
instead decreases. That is why, in spite of its simplicity, the
demonstration of the Deutsch equation in laboratory is not a trivial
task. One needs to find a two-particle scattering event, where the
density matrix of one of the particles remains the same after the
scattering.

\section{Eigenvalue Formulation}

Let CR and CTC registers be single qubits. For a pure state
$\rho_{CR}=|\psi\rangle\langle\psi|$, without loss of generality
$\rho_{CR}$ can be assumed to be given by $|0\rangle\langle 0|$
since $\tilde{U}=U\left(V\otimes I\right)$, where
$|\psi\rangle=V|0\rangle$, is also unitary. For
$\rho_{CR}=|0\rangle\langle 0|$ and
\begin{eqnarray}
\rho_{CTC}= \left(\begin{array}{cc}
\rho_{11} & \rho_{12} \\
    \rho_{21}& \rho_{22} \\
\end{array}\right),
\end{eqnarray}
inserting matrix elements of $\rho_{CR}$ and $\rho_{CTC}$, and
comparing the entries of its left and right hand sides,
Eq.~(\ref{de}) can be written as $M\vec{v}=\vec{v}$ where $M$ is
given in terms of the matrix elements $u_{ij}$ of unitary operator
$\tilde{U}$, and their complex conjugates $u_{ij}^{*}$, as
\begin{eqnarray}
M= \left(\begin{array}{cccc}
u_{11}u_{11}^*+u_{31}u_{31}^* & u_{11}u_{12}^*+u_{31}u_{32}^* & u_{12}u_{11}^*+u_{32}u_{31}^* & u_{12}u_{12}^*+u_{32}u_{32}^*  \\
u_{11}u_{21}^*+u_{31}u_{41}^* & u_{11}u_{22}^*+u_{31}u_{42}^* & u_{12}u_{21}^*+u_{32}u_{41}^* & u_{12}u_{22}^*+u_{32}u_{42}^*  \\
u_{21}u_{11}^*+u_{41}u_{31}^* & u_{11}u_{12}^*+u_{41}u_{32}^* & u_{22}u_{11}^*+u_{42}u_{31}^* & u_{22}u_{12}^*+u_{42}u_{32}^*  \\
u_{21}u_{21}^*+u_{41}u_{41}^* & u_{21}u_{22}^*+u_{41}u_{42}^* & u_{22}u_{21}^*+u_{42}u_{41}^* & u_{22}u_{22}^*+u_{42}u_{42}^*  \\
\end{array}\right),
\end{eqnarray}
and $\vec{v}=\left(\rho_{11}, \rho_{12}, \rho_{21},
\rho_{22}\right)^T$. Here, $\vec{v}$ is normalized so as to satisfy
$\rho_{11}+\rho_{22}=1$. Therefore, the Deutsch equation can be
interpreted as a diagonalization problem $M\vec{v}=\lambda\vec{v}$,
where one looks for $\lambda=1$ eigenvalues and vectors $\vec{v}$
whose entries define a valid density matrix.

Although Deutsch gave a general proof for existence of a solution,
it is worth rederiving the same result using the current formalism
due to its simplicity. Using the unitarity of $\tilde{U}$, it is
seen that the matrix $M$ is of the form
\begin{eqnarray}
M= \left(\begin{array}{cccc}
a & b & b^* & c \\
d & e & f^* & g \\
d^* & f & e^* & g^* \\
1-a & -b & -b^* & 1-c \\
\end{array}\right).
\end{eqnarray}
Adding the last row of the matrix $M-\lambda I$ ($I$ being the
identity matrix) to its first row, one can immediately observe that
$1-\lambda$ is always a factor of the characteristic equation
$\det\left(M-\lambda I\right)$. Therefore, there is at least one
solution. If $\lambda=1$ is two-fold degenerate, any convex linear
combination, $\alpha\vec{v}_1+\left(1-\alpha\right)\vec{v}_2$ with
$\alpha\in[0,1]$, of the corresponding eigenvectors $\vec{v}_1$ and
$\vec{v}_2$ is also a solution. As shall be seen below, there can be
cases where $\vec{v}_2$ does not correspond a valid density matrix,
while its convex linear combinations with $\vec{v}_1$ are proper
solutions.

A practical way to construct $M$ is to write it as $M=A_{11}\otimes
A_{11}^*+A_{21}\otimes A_{21}^*$ where $A_{11}$ and $A_{21}$ are
$2\times 2$ matrices in
\begin{eqnarray}
\tilde{U}= \left(\begin{array}{cc}
A_{11} & A_{12} \\
A_{21} & A_{22} \\
\end{array}\right),
\label{unit}
\end{eqnarray}
and $A_{ij}^{*}$ denotes complex conjugation of $A_{ij}$.

Generalization to $n$ CR qubits near a CTC qubit problem is
straightforward. In this case,
\begin{equation}
M=\sum_{i=1}^{2^{n}}A_{i1}\otimes A_{i1}^{*},
\end{equation}
where $A_{ij}$ are $2\times 2$ matrices making up the $2^{n+1}\times
2^{n+1}$ unitary matrix $\tilde{U}$ as in Eq.~(\ref{unit}). Even
though the eigenvalue formalism has been presented for pure a
$\rho_{CR}$, it can easily be generalized to mixed states. In this
case $M$ matrix is slightly more complicated, and it contains all
four entries of $\rho_{CR}$.

\section{Disappearance of Entanglement Near a CTC}

An interesting problem involving entanglement near a CTC is the
behavior of an EPR pair. In order to illustrate Deutsch's model,
Bennett \emph{et al.}, considered half of a maximally entangled
state, i.e., an EPR pair
$\frac{1}{\sqrt{2}}\left(|00\rangle+|11\rangle\right)$, and put it
into a CTC \cite{Bennett}. Using both the single and multiple
universe pictures, the authors showed that the joint state at any
time after the interaction between CR and CTC systems is a product
state. In the eigenvalue formulation introduced in Sec. IV, the
total unitary interaction $\tilde{U}$ involving three qubits (i.e.,
two CR qubits making up the EPR pair and one CTC qubit) can be
written as $\tilde{U}=\left(I_{2}\otimes E\right)\left(CNOT\otimes
I_{2}\right)\left(H\otimes I_{4}\right)$ where $H$ and $CNOT$ denote
the Hadamard and controlled-NOT operations, respectively. The EPR
state is obtained by acting the operator $\left(CNOT\otimes
I_{2}\right)\left(H\otimes I_{4}\right)$ on the initial CR two qubit
product state $|00\rangle$, where $I_{2}$ and $I_{4}$ denote
$2\times2$ and $4\times4$ unit matrices, respectively. Here,
\begin{eqnarray}
E= \left(\begin{array}{cccc}
1 & 0 & 0 & 0 \\
0 & 0 & 1 & 0 \\
0 & 1 & 0 & 0 \\
0 & 0 & 0 & 1 \\
\end{array}\right),
\end{eqnarray}
denotes the exchange operation, where a qubit emerges from the CTC
and half of the EPR pair is put into the CTC.

Diagonalization of the corresponding $M$ matrix gives the
eigenvalues $\lambda_{1}=1$, and
$\lambda_{2}=\lambda_{3}=\lambda_{4}=0$. Hence, the solution is
unique, and the reduced density matrices for CR and CTC registers
are $I_{4}/4$ and $I_{2}/2$, respectively. Therefore, two CR qubits
are not entangled anymore. Since the reduced density matrix of the
CR qubits is found to be mixed after the interaction, this is an
expected result. It is known that according to the monogamy property
of entanglement, if two qubits are maximally entangled, they cannot
be entangled with a third qubit \cite{Coffman}.

\section{Discontinuous Evolutions Near CTCs}

DeJonghe \emph{et al.}, demonstrated that Deutsch's equation can
lead to discontinuities in the evolution of the CR systems
\cite{DeJonghe}. The authors consider two CR qubits interacting with
a CTC qubit via the unitary evolution
\begin{eqnarray}
\begin{array}{r}
U= |000\rangle\langle100|+|100\rangle\langle000|+|010\rangle\langle011|+|011\rangle\langle010|\\
    |101\rangle\langle110|+|110\rangle\langle101|+|001\rangle\langle001|+|111\rangle\langle111|.
\end{array}
\end{eqnarray}
It is also assumed that the density matrix of CR qubits before the
interaction is of the form $\rho_{CR}=\rho_{1}\otimes\rho_{2}$. For
three distinct initial states
($\rho_{CR}^{A},\rho_{CR}^{B},\rho_{CR}^{C}$) which are
infinitesimally close to each other, it has been shown that there is
no choice of $\rho_{CTC}$ which is continuous in the vicinity of
$\rho_{CR}^{B}$. Since the CR and CTC density matrices are obtained
from the same matrix by partial trace operations, the discontinuity
in the CTC state implies the behavior of the  CR state. In the
current formalism, starting again from CR state $|00\rangle$, the
three pure states $\rho_{CR}^{A},\rho_{CR}^{B},\rho_{CR}^{C}$
correspond to the following $M$ matrices
\begin{eqnarray}
M^{A}= \left(\begin{array}{cccc}
1-\epsilon & 0 & 0 & \epsilon \\
0 & 0 & \epsilon & 0 \\
0 & \epsilon & 0 & 0 \\
\epsilon & 0 & 0 & 1-\epsilon \\
\end{array}\right),
\end{eqnarray}
\begin{eqnarray}
M^{B}= \left(\begin{array}{cccc}
1 & 0 & 0 & 0 \\
0 & 0 & 0 & 0 \\
0 & 0 & 0 & 0 \\
0 & 0 & 0 & 1 \\
\end{array}\right),
\end{eqnarray}
\begin{eqnarray}
M^{C}= \left(\begin{array}{cccc}
1 & 0 & 0 & \epsilon \\
0 & \sqrt{\epsilon(1-\epsilon)} & 0 & 0 \\
0 & 0 & \sqrt{\epsilon(1-\epsilon)} & 0 \\
0 & 0 & 0 & 1-\epsilon \\
\end{array}\right)
\end{eqnarray}
where $\epsilon\in[0,1]$. Eigenvalues for three cases are given by
$\lambda_{1}^{A}=1,\lambda_{2}^{A}=1-2\epsilon,
\lambda_{3}^{A}=-\epsilon, \lambda_{4}^{A}=\epsilon$ for initial
state $\rho_{CR}^{A}$, $\lambda_{1}^{B}=\lambda_{2}^{B}=1,
\lambda_{3}^{B}=\lambda_{4}^{B}=0$ for $\rho_{CR}^{B}$, and
$\lambda_{1}^{C}=1,\lambda_{2}^{C}=1-\epsilon,
\lambda_{3}^{C}=\lambda_{4}^{C}=\sqrt{\epsilon(1-\epsilon)}$ for
$\rho_{CR}^{C}$. Any eigenvector with eigenvalue 1 is a solution of
the Deutsch equation. Degeneracy of $\lambda_{1}^{B}$ and
$\lambda_{2}^{B}$ indicate that for initial state $\rho_{CR}^{B}$
there are infinitely many solutions which can be obtained by taking
convex linear combinations of the two degenerate states. CTC state
solutions, all of which are independent of $\epsilon$, for the three
cases are given by
\begin{eqnarray}
\rho_{CTC}^{A}= \left(\begin{array}{cc}
1/2 & 0 \\
0 & 1/2 \\
\end{array}\right),
\end{eqnarray}
\begin{eqnarray}
\rho_{CTC}^{B}= \left(\begin{array}{cc}
\beta & 0 \\
0 & 1-\beta \\
\end{array}\right),  \,\mbox{for} \,\, \beta\in[0,1],
\label{rhob}
\end{eqnarray}
\begin{eqnarray}
\rho_{CTC}^{C}= \left(\begin{array}{cc}
1 & 0 \\
0 & 0\\
\end{array}\right).
\end{eqnarray}
Density matrices $\rho_{CTC}^{A}$ and $\rho_{CTC}^{C}$ can be
obtained from $\rho_{CTC}^{B}$ by choosing $\beta=1/2$ and
$\beta=1$, respectively. As $\epsilon\rightarrow0$, all three
initial states approach each other, but $\rho_{CTC}^{A}$ and
$\rho_{CTC}^{C}$, being independent of $\epsilon$, are always
different. Solving the Deutsch equation by a different method,
DeJonghe \emph{et al.} concluded that there is a discontinuity near
$\rho_{CTC}^{B}$ for the unitary interaction given above. Therefore,
for finite $\epsilon$ values, the eigenvalue formulation reproduces
their result. However, as $\epsilon\rightarrow0$, eigenvalues
$\lambda_{2}^{A}, \lambda_{2}^{C}\rightarrow1$, and hence the
corresponding density matrices satisfy the Deutsch equation. At this
limit, all three initial states lead to the same solution set, given
by Eq.~(\ref{rhob}), for $\rho_{CTC}$. More explicitly, approximate
solutions obtained by taking linear combinations of eigenvectors
with eigenvalues $\lambda_{1}$ and $\lambda_{2}$ for the initial
states $\rho_{CR}^{A}$ and $\rho_{CR}^{C}$ can be written as
\begin{eqnarray}
\rho_{CTC}^{A}= \left(\begin{array}{cc}
1/2-\alpha & 0 \\
0 & 1/2+\alpha\\
\end{array}\right), \,\mbox{for} \,\, \alpha\in[-1/2,1/2],
\end{eqnarray}
and
\begin{eqnarray}
\rho_{CTC}^{C}= \left(\begin{array}{cc}
1-\gamma & 0 \\
0 & \gamma\\
\end{array}\right), \,\mbox{for} \,\, \gamma\in[0,1].
\end{eqnarray}
For both $\rho_{CR}^{A}$ and $\rho_{CR}^{C}$, $\lambda_{2}$ solution
is an $\epsilon$ independent vector $\vec{v}_{2}=\left(-1, 0, 0,
1\right)^T$ which gives a traceless $2\times2$ matrix. Even though
this solution cannot be associated with a density matrix, its linear
combinations with $\vec{v}_{1}$ (eigenvector with eigenvalue
$\lambda_{1}=1$), lead to proper vectors. Solution set for all three
cases at $\epsilon\rightarrow0$ is the same. For $\rho_{CR}^{B}$,
the expression in Eq.~(\ref{rhob}) is a solution for all
$\beta\in[0,1]$. However, in case of $\rho_{CR}^{A}$ and
$\rho_{CR}^{C}$, the only $\epsilon$-independent and exact solutions
are $\alpha=0$ and $\gamma=0$, respectively.

\section{Conclusion}

The consistency condition proposed by Deutsch to avoid paradoxes
near a CTC has been transformed to an eigenvalue equation. The
proposed approach is a systematic method that solves the Deutsch
equation. Two problems, namely an EPR pair near a CTC and the
discontinuous evolution of the CR-CTC system, have been reexamined.
Even though the eigenvalue formulation has been demonstrated for
initially pure CR states, it can easily be generalized to the mixed
case.

\section{Acknowledgment}
This work has been partially supported by the Scientific and
Technological Research Council of Turkey (T\"{U}B\.ITAK) under
Grants 107T530 and 111T232. The author would like to thank A.~Aliev,
\"{O}.~Er\c{c}etin, G.~Karpat, and C.~Sa\c{c}l{\i}o\u{g}lu for
helpful discussions.

\end{document}